\documentclass[preprint2]{emulateapj}

\shorttitle{Recombining Plasma in IC~443}
\shortauthors{Yamaguchi et al.}

\begin{document}

\title{Discovery of Strong Radiative Recombination Continua from \\
The Supernova Remnant IC~443 with Suzaku}

\author{H. Yamaguchi\altaffilmark{1}}
\email{hiroya@crab.riken.jp}

\author{M. Ozawa\altaffilmark{2}}
\author{K. Koyama\altaffilmark{2}}
\author{K. Masai\altaffilmark{3}}
\author{J. S. Hiraga\altaffilmark{1}}
\author{M. Ozaki\altaffilmark{4}}
\author{D. Yonetoku\altaffilmark{5}}


\altaffiltext{1}{RIKEN (The Institute of Physical and Chemical Research), 
  2-1 Hirosawa, Wako, Saitama 351-0198, Japan}
\altaffiltext{2}{Department of Physics, Kyoto University, 
  Kitashirakawa-oiwake-cho, Sakyo-ku, Kyoto 606-8502, Japan}
\altaffiltext{3}{Department of Physics, Tokyo Metropolitan University, 
  1-1 Minami-Osawa, Hachioji, Tokyo 192-0397, Japan}
\altaffiltext{4}{Institute of Space and Astronautical Science, JAXA, 
  3-1-1 Yoshinodai, Sagamihara, Kanagawa 229-8510, Japan}
\altaffiltext{5}{Department of Physics, Kanazawa University, 
  Kakuma-machi, Kanazawa, Ishikawa 920-1192, Japan}

\begin{abstract}

We present the {\it Suzaku} spectroscopic study of the Galactic middle-aged 
supernova remnant (SNR) IC~443. The X-ray spectrum in the 1.75--6.0~keV band 
is described by an optically-thin thermal plasma with the electron temperature 
of $\sim 0.6$~keV and several additional Lyman lines. We robustly detect, 
for the first time, strong radiative recombination 
continua (RRC) of H-like Si and S around at 2.7 and 3.5~keV. The ionization 
temperatures of Si and S determined from the intensity ratios of the RRC to 
He-like K$\alpha$ line are $\sim 1.0$~keV and $\sim 1.2$~keV, respectively. 
We thus find firm evidence for an extremely-overionized (recombining) plasma. 
As the origin of the overionization, a thermal conduction scenario argued 
in previous work is not favored in our new results. We propose that 
the highly-ionized gas were made at the initial phase of the SNR evolution 
in dense regions around a massive progenitor, and the low electron 
temperature is due to a rapid cooling by an adiabatic expansion. 

\end{abstract}

\keywords{ISM: individual (IC~443) --- supernova remnants 
--- radiation mechanisms: thermal 
--- X-rays: ISM}

\section{Introduction}
\label{sec:introduction}

IC~443 (G189.1+3.0), a Galactic supernova remnant (SNR) at a distance 
of 1.5~kpc (Welsh \& Sallmen 2003), is located near the Gem OB1 association 
and a dense giant molecular cloud (Cornett et al.\ 1977) with OH maser 
emission (Claussen et al.\ 1997). These facts strongly suggest that the 
remnant originated from a collapse of a massive progenitor. 
Braun \& Strom (1986) proposed that the shock wave has expanded into the 
pre-existing wind-blown bubble shell. This was confirmed by a kinematical 
study of the optical filaments (Meaburn et al.\ 1990). 
A comprehensive X-ray study of IC~443 was first made with the 
{\it Einstein} and {\it HEAO-A2} satellites (Petre et al.\ 1988). 
They estimated the SNR age to be $\sim$3000~yr. 
Recently, Troja et al.\ (2008) derived the age of $\sim$4000~yr from 
the morphologies of the shocked ejecta and interstellar medium (ISM) 
revealed by {\it XMM-Newton}. Thus, IC~443 is a middle-aged SNR.

Using {\it ASCA}, Kawasaki et al.\ (2002) found that the ionization 
degrees of Si and S were significantly higher than those expected from 
the electron temperature of the bremsstrahlung continuum. 
Therefore, it was argued that the plasma is in overionization 
($kT_z$ $>$ $kT_e$, where $kT_z$ and $kT_e$ are ionization and electron 
temperatures, respectively). 
On the other hand, Troja et al.\ (2008) found that the plasma is in 
collisional ionization equilibrium (CIE), or the overionization is 
only marginal based on {\it XMM-Newton} data. 
These two controversial results hinge upon small differences, 
if any, between the estimated electron and ionization temperatures.

In this letter, we investigate whether the overionized plasma is really 
present or not, by utilizing the superior spectral capabilities for 
diffuse sources of X-ray Imaging Spectrometers (XIS: Koyama et al.\ 2007) 
aboard the {\it Suzaku} satellite (Mitsuda et al.\ 2007). 
If present, we will study the plasma condition quantitatively 
to discuss the possible origin of the overionization.

\section{Observation and Data Reduction}
\label{sec:observation}

The northern part of IC~443 was observed with {\it Suzaku} on 2007 March 6 
(Observation ID = 501006010). Three XISs located at the foci of the 
independent X-Ray Telescopes (XRT: Serlemitsos et al.\ 2007) were 
operating\footnote{{\it Suzaku} carried four XISs, but one is now out of 
operation due to a damage possibly by an impact of a micro-meteorite}. 
Two of the XISs are Front-Illuminated (FI) CCDs and the other is 
a Back-Illuminated (BI) CCD. 
The XIS was operated in the normal full-frame clocking mode with 
a spaced-row charge injection technique (Uchiyama et al.\ 2009) 
during the observation. 
For the data reduction and spectral analysis, we used the HEADAS software 
package of version 6.5 and XSPEC version 11.3.2, respectively. The XIS data 
of revision 2.0 were employed, but reprocessed using the \texttt{xispi} 
software and the latest calibration database released on 2009 February 3. 
After the screening with the standard criteria\footnote{http://heasarc.nasa.gov/docs/suzaku/processing/criteria\_xis.html}, 
an effective exposure of $\sim$42~ksec was obtained.

\section{Analysis and Results}
\label{sec:result}

\begin{figure}[t]
  \includegraphics[scale=.50]{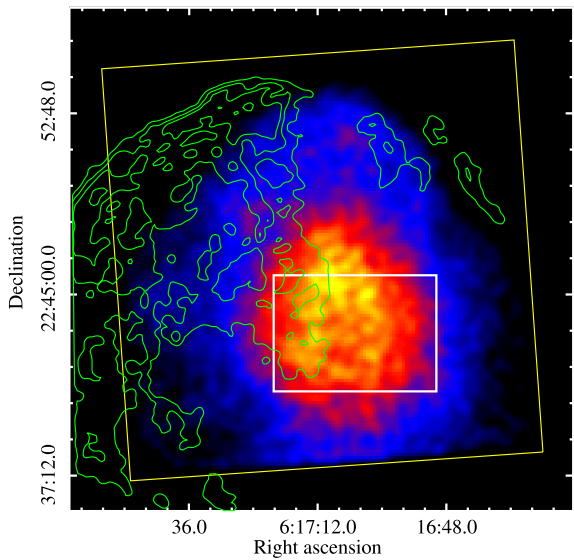}
\caption{Vignetting-corrected XIS image of the northern part of IC~443 
  in the 1.75--3.0~keV band, shown on an intensity color scale. 
  The coordinates (R.A.\ and Dec.) refer to epoch J2000.0. 
  The data from the three active XISs are combined and smoothed with 
  a Gaussian kernel of $\sigma = 25''$. 
  The yellow square and the white rectangle indicate the XIS field of view 
  and the region used in our spectral analysis, respectively. 
  The optical Digitized Sky Survey image is overplotted in contour.
\label{fig:image}}
\end{figure}

\begin{figure}[t]
\includegraphics[scale=.35]{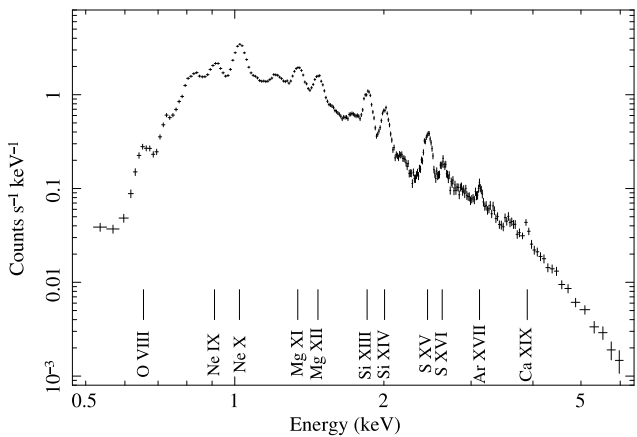}
\caption{Full-band XIS-FI spectrum where the NXB is subtracted. 
  The energies of prominent K$\alpha$ emission lines from 
  specific elements are labeled in the panel. 
  \label{fig:full}}
\end{figure}

\begin{figure}[t]
\includegraphics[scale=.35]{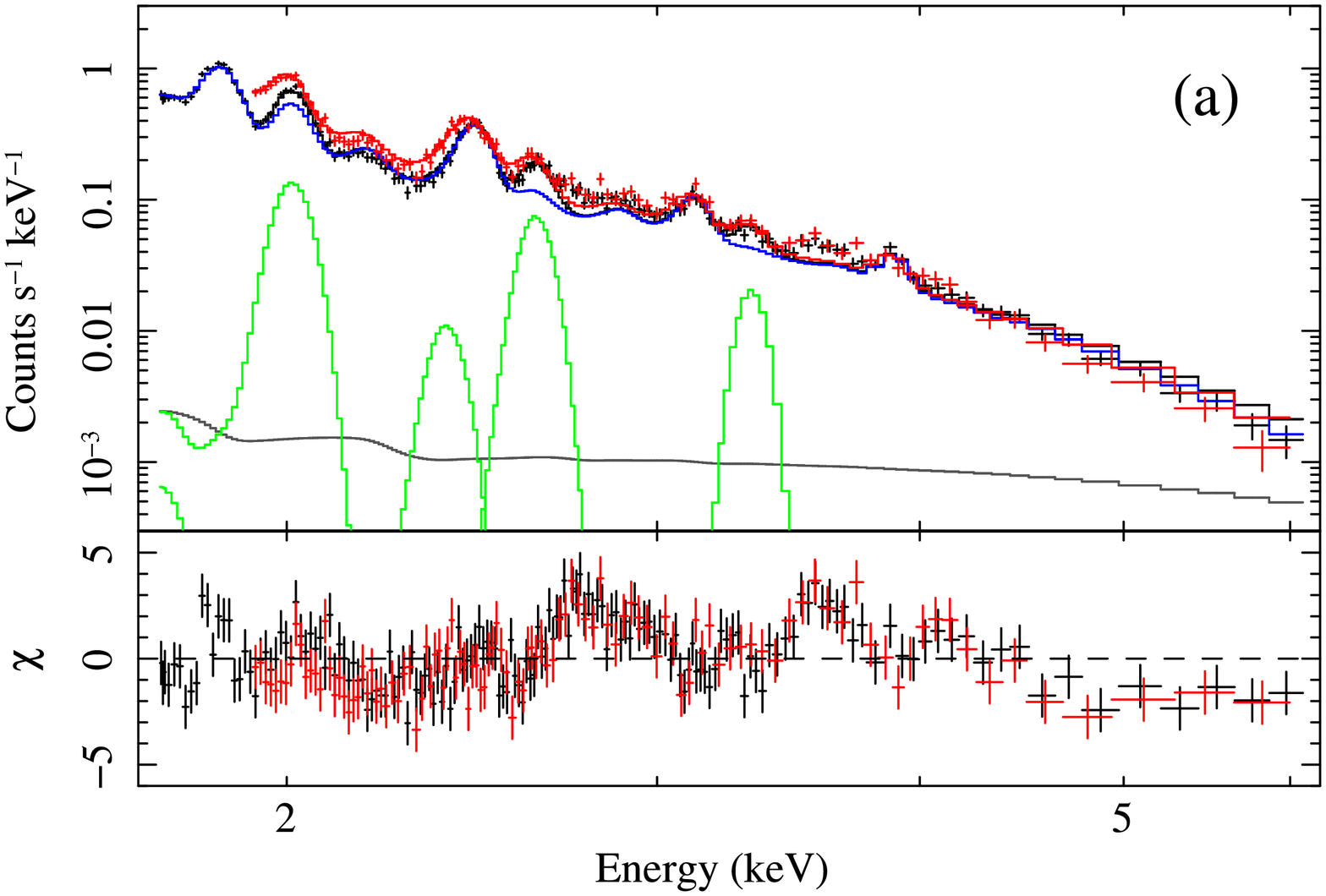}
\includegraphics[scale=.35]{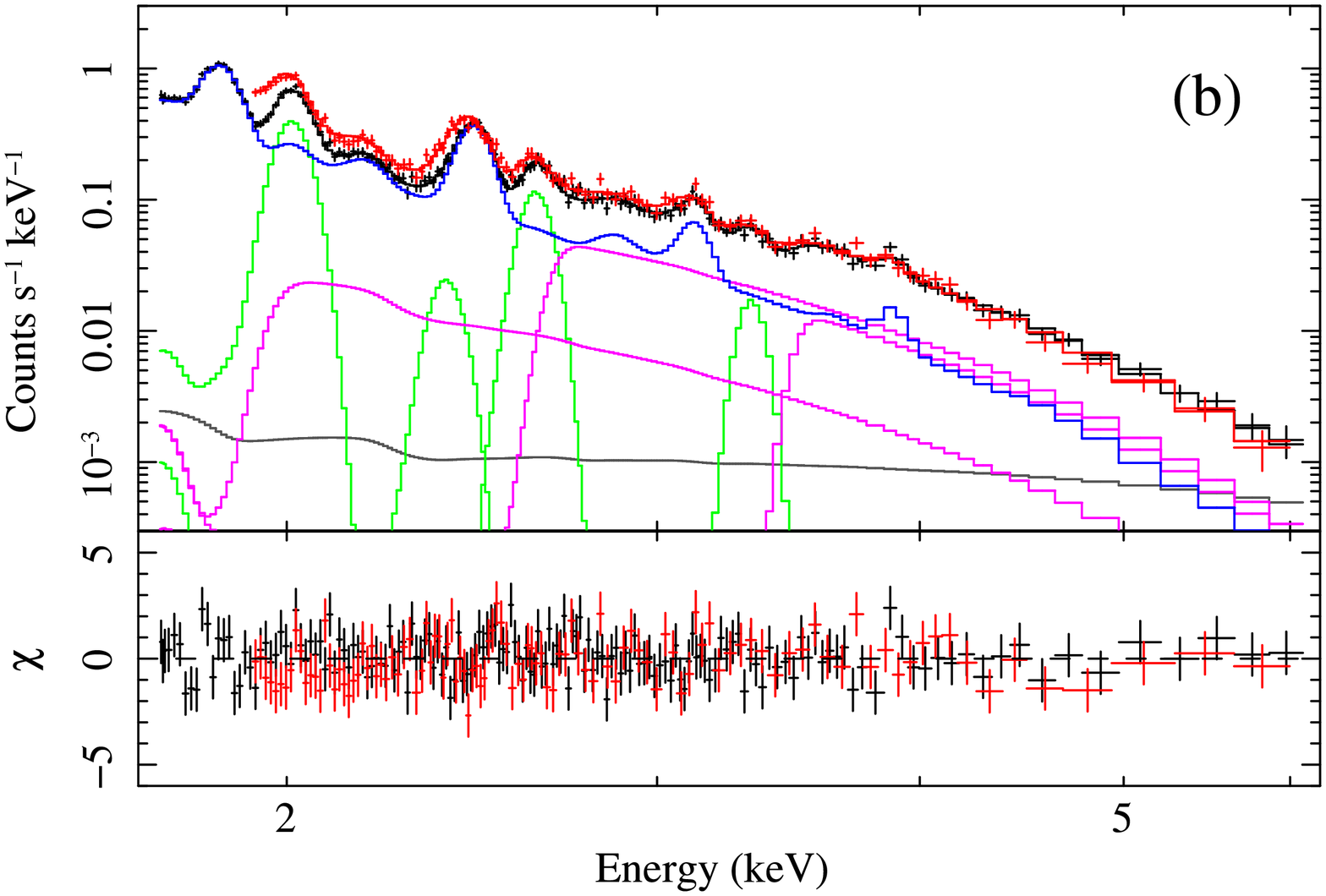}
\caption{(a) XIS spectrum in the 1.75--6.0~keV band. 
  Black and red represent FI and BI, respectively. 
  Individual components of the best-fit model for the FI data are shown 
  with solid colored lines: blue, green, and gray are the VAPEC, 
  Gaussians (Si-Ly$\alpha$, Si-Ly$\beta$, S-Ly$\alpha$, and Ar-Ly$\alpha$), 
  and CXB, respectively. 
  The lower panel shows the residual from the best-fit model.
  Two hump-like features are clearly found around the energies of 
  $\sim$2.7~keV and $\sim$3.5~keV. \
  (b) Same spectrum as (a), but for a fit with RRC components 
  of H-like Mg, Si, and S (magenta lines). 
  The residuals seen in (a) are disappeared. \
  \label{fig:fit}}
\end{figure}

Figure~\ref{fig:image} shows the vignetting-corrected XIS image in 
1.75--3.0~keV, the energy band including the major lines of K-shell 
emissions from Si and S. We extracted the spectrum from the brightest 
region, the rectangular with an angular size of $7' \times 5'$ 
(see Fig.~\ref{fig:image}). This approximately corresponds to the ``Center'' 
region of Kawasaki et al.\ (2002).
The XIS spectra of the two FIs in the entire energy range 
were made by subtracting the non X-ray background (NXB) 
constructed with the \texttt{xisnxbgen} software. 
The spectra were merged to improve the statistics, 
because the response functions are almost identical each other. 
Figure~\ref{fig:full} shows the resultant spectrum. 
We can see several prominent lines of K$\alpha$ emission from He- and 
H-like ions (hereafter, He$\alpha$ and Ly$\alpha$). 
The centroids of the Ly$\alpha$ lines were measured with a Gaussian line
model, and compared with the canonical values of the Astrophysical Plasma 
Emission Database (APED: Smith et al.\ 2001).
The averaged center energy difference was +4~eV. 
Therefore, we added a 4~eV offset in the FI spectrum. 
The BI spectrum was made with the same procedure as the FIs, 
but an offset of $-10$~eV was added to correct the energy scale.
In order to examine the ionization states of Si and S, we hereafter focus 
on the spectrum in the energy range above 1.75~keV (Fig.~\ref{fig:fit}). 
Detailed studies including the lower energies will be reported in 
a separate paper.

We first fitted the spectrum with a model of a thin-thermal plasma 
in CIE state (a VAPEC model). 
The abundances (Anders \& Grevesse 1989) of Si, S, and Ar were free 
parameters, while the Ca abundance was tied to Ar. 
Interstellar extinction was fixed to a hydrogen column density of 
$N_{\rm H}$ = $7\times 10^{21}$~cm$^{-2}$ with the solar elemental 
abundances, following Kawasaki et al.\ (2002) and Troja et al.\ (2008). 
The cosmic X-ray background (CXB) spectrum was approximated by a power-law 
model with photon index of $\Gamma$ = 1.412 and the surface brightness in 
the 2--10~keV band of $6.4 \times 10^{-8}$~erg~cm$^{-2}$~s$^{-1}$~sr$^{-1}$ 
(Kushino et al.\ 2002).
Since IC~443 is located in the anti-Galactic center direction, 
contribution of the Galactic ridge X-ray emission was ignored.
In the initial fit, we found a significant inconsistency between 
the FI and BI data around the energy of neutral Si K-edge (1.84~keV). 
This is due to the well-known calibration issue of the XIS. 
Since the calibration for the FI CCDs is currently far better than the BI, 
we decided to ignore the energy band below 1.9~keV in the BI spectrum. 
This fit leaved further large residuals, in both spectra, 
at the energies of S and Ar Ly$\alpha$ lines, 
and hence was rejected with the $\chi^2$/dof of 935/270.

We, therefore, applied a model of one-temperature VAPEC (CIE plasma) 
plus narrow Gaussian lines at 2006, 2623, and 3323~eV, 
the Ly$\alpha$ energies of the Si, S, and Ar, hydrogenic ions, respectively. 
This process is essentially the same as Kawasaki et al.\ (2002). 
Ly$\beta$ lines of the same elements were also added, but they were 
not significant except for Si (2377~eV). 
The result is shown in Figure~\ref{fig:fit}a, with the best-fit CIE 
plasma temperature of $kT_e$ $\sim$ 0.94~keV. 
Although the $\chi ^2$/dof value was significantly reduced to 657/266, 
the model was still unacceptable. In fact, apparent hump-like residuals are
found around $\sim$2.7~keV and $\sim$3.5~keV, in addition to the systematic 
model excess in the continuum at the energies above $\sim$4.5~keV.

We checked systematic error due to the CXB fluctuation by allowing the
CXB intensity as a free  parameter. The fit, however, did not improve at all. 
We also tried two-component VAPEC models with independent temperatures and 
abundances, but these were rejected with $\chi ^2$/dof = 737/265. 
The results were essentially the same as figure~\ref{fig:fit}a; the hump-like 
residuals remained at $\sim$2.7~keV and $\sim$3.5~keV. 
No additional CIE component nor non-equilibrium ionization (NEI) plasma 
model removed the hump-like residuals with significant improvement of 
the $\chi^2$ value.

\begin{table}[t]
\begin{center}
\caption{Best-fit spectral parameters 
  \label{tab:fit}}
\begin{tabular}{llcc}
  \tableline\tableline
  Component & \multicolumn{2}{c}{Parameter} & Value  \\ 
  \tableline
  CIE (VAPEC) & \multicolumn{2}{l}{$kT_e$ (keV)} & 0.61 (0.59--0.64) \\
  ~ & \multicolumn{2}{l}{$Z_{\rm Si}$ (solar)} & 0.82 (0.78--0.85) \\
  ~ & \multicolumn{2}{l}{$Z_{\rm S}$ (solar)} & 1.7 (1.6--1.8) \\
  ~ & \multicolumn{2}{l}{$Z_{\rm Ar}$ (solar)} & 2.5 (2.2--2.8) \\
  ~ & \multicolumn{2}{l}{VEM\tablenotemark{a} ($10^{12}$~cm$^{-5}$)} & 
     6.4 (6.3--6.6) \\
  \tableline
  \multicolumn{4}{c}{Additional components} \\
  \tableline
  ~ & ~ & $E$ (keV)\tablenotemark{b} &  Flux\tablenotemark{c} \\
  \tableline
  Line  & Si Ly$\alpha$ & 2.006 & 2.8 (2.7--2.9)  \\
  ~     & Si Ly$\beta$  & 2.377 & 0.21 (0.14--0.29)  \\
  ~     & S Ly$\alpha$  & 2.623 & 0.84 (0.79--0.90)  \\
  ~     & Ar Ly$\alpha$ & 3.323 & 0.11 (0.085--0.14)  \\
  RRC & H-like Mg & 1.958 & 1.2 (1.0--1.5)  \\
  ~   & H-like Si  & 2.666 & 2.2 (2.0--2.3)  \\
  ~   & H-like S  & 3.482 & 0.46 (0.41--0.51)  \\
  \tableline
\end{tabular}
\end{center}
\tablecomments{
  The uncertainties in the parentheses are the 90\% confidence range. 
  $^{a}$Volume emission measure, $\int n_e n_p~dV/(4\pi D^2)$, 
  where $n_e$, $n_p$, $V$, and $D$ are the electron and proton densities 
  (cm$^{-3}$), the emitting volume (cm$^3$), and the distance to 
  the source (cm), respectively. \ 
  $^{b}$The fixed energy values of the line center or the K-edge.\ 
  $^{c}$Total flux in the unit of $10^{-4}$ photon cm$^{-2}$ s$^{-1}$.\\}
\end{table}

At the energies of the humps, no emission line candidate from an abundant 
element is found. However, the energies are consistent with the K-shell 
binding potentials ($I_z$) of the H-like Si (2666~eV) and S (3482~eV). 
Therefore, the humps are likely due to the free-bound transitions to the 
K-shell of the H-like Si and S. 
A formula for the spectrum of radiative recombination continuum (RRC) is 
found in equation~(21) of Smith \& Brickhouse~(2002)\footnote{http://cxc.harvard.edu/atomdb/physics/plasma/plasma.html}.
When the electron temperature is much lower than the K-edge energy 
($kT_e$ $\ll$ $I_z$), this formula is approximated as; 
\begin{equation}
\frac{dP}{dE}(E_{\gamma}) \propto 
{\rm exp}\left( -\frac{E_{\gamma} - I_z}{kT_e}\right), 
~~~ {\rm for}~ E_{\gamma} \geq I_z~.
\label{eq:rrc}
\end{equation}
Thus, the width of the RRC structure depends on the electron temperature. 
We added the RRC of Equation~\ref{eq:rrc} for H-like\footnote{``H-like RRC'' 
refers to the free-bound emission due to electron captures 
by fully-ionized ions into ground state of H-like ions} Si and S. 
The $kT_e$ values of the RRC were linked to that of the VAPEC component. 
Then, the fit was dramatically improved to an acceptable $\chi ^2$/dof 
of 290/264. 
Although the hump-like residuals were completely removed 
by this fit, we further added the RRC model of H-like Mg. 
This step is reasonable because Mg is more abundant than Si and S in 
the solar abundance ratio of Anders \& Grevesse (1989) and the K-edge 
energy of H-like Mg ($I_z$ = 1958~eV) falls into the analyzed band. 
The $\chi ^2$/dof value was significantly reduced to 267/263, 
which gives an $F$-test probability of $\sim 3 \times 10^{-6}$. 
The best-fit parameters and model are given in Table~\ref{tab:fit} and 
Figure~\ref{fig:fit}b, respectively. 
The spectrum was also fitted with the independent electron temperatures 
for the VAPEC and RRC components, but the values are consistent with 
each other within their statistical uncertainties.

\section{Discussion and Conclusion}
\label{sec:discussion}

\begin{figure}[t]
\includegraphics[scale=.35]{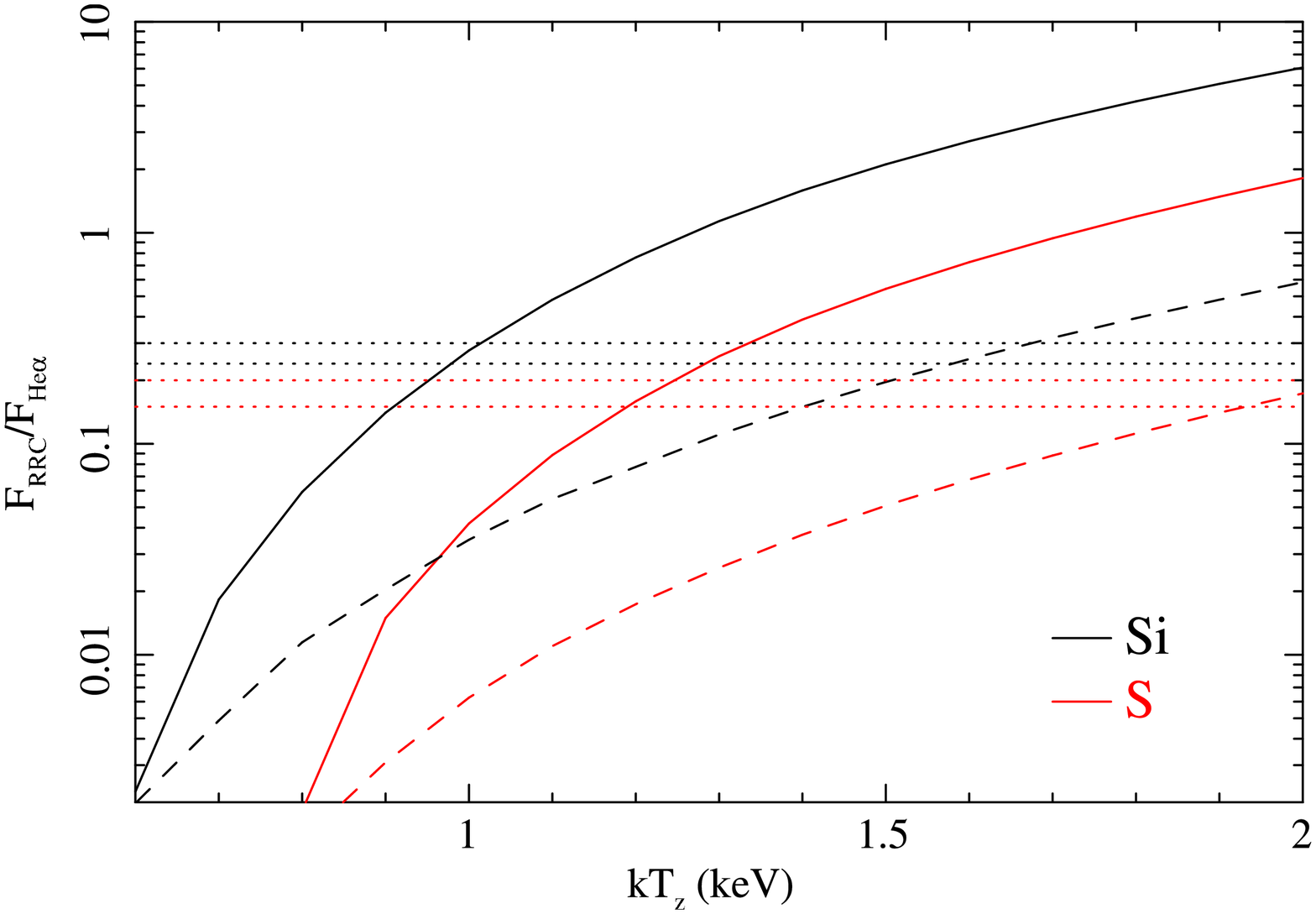}
\caption{Emissivity ratios of H-like RRC to He-like K$\alpha$ lines 
  as a function of ionization temperature ($kT_z$), 
  predicted by the plasma radiation code of Masai (1994) 
  for recombining plasma with electron temperature ($kT_e$) of 0.6~keV. 
  Black and red solid lines represent Si and S, respectively. 
  For comparison, the same ratios for CIE ($kT_z$ = $kT_e$) plasma 
  (the APEC model: Smith 2001) are indicated by dashed lines. 
  The horizontal dotted lines represent the 90\% upper and lower limits 
  of the observed values. 
  \label{fig:ratio}}
\end{figure}

We have found that the 1.75--6.0~keV spectrum cannot be represented 
with CIE nor NEI plasma alone, but need the additional fluxes of 
the Lyman lines and the H-like RRC of Si and S (and possibly Mg). 
This is the first detection of clear RRC emissions in an SNR. 
In the following, we quantitatively discuss the implications of 
our spectral results.

From the best-fit model in Table~\ref{tab:fit}, the flux ratios of H-like 
RRC to He-like K$\alpha$ (He$\alpha$) line ($F_{\rm RRC}/F_{{\rm He}\alpha}$) 
are given to be 0.28 (0.24--0.30) and 0.18 (0.15--0.20) for Si and S, 
respectively. These are compared, in Figure~\ref{fig:ratio}, with the 
modeled emissivity ratios by the plasma radiation code of Masai (1994) 
for the electron temperature of 0.6~keV. 
We find that the large observed ratios of $F_{\rm RRC}/F_{{\rm He}\alpha}$ 
are significantly above those in the CIE case ($kT_z$ = $kT_e$), 
but can be reproduced in the overionization case ($kT_z$ $>$ $kT_e$). 
The ionization temperatures of Si and S are determined to be 
$\sim 1.0$~keV and $\sim 1.2$~keV, respectively. This is, therefore, 
the firm evidence of the overionized (recombining) plasma.

The overionization claim for IC~443 was first argued by Kawasaki et al.\ 
(2002). Using {\it ASCA} data, they derived $kT_z$ to be $\sim 1.5$~keV 
from the S Ly$\alpha$/He$\alpha$ flux ratio compared with the predicted 
emissivity ratio in the CIE plasma code. 
Troja et al.\ (2008) adopted the same analysis procedure to the 
{\it XMM-Newton} spectrum, and claimed that $kT_z$ obtained from the 
line flux ratio was nearly same as the bremsstrahlung temperature ($kT_e$). 
However, since the RRC process is accompanied with electron captures to the 
excited levels, as given in Equation~\ref{eq:rate} below, the resulting 
cascade decay to the ground state contributes the line emission. 
In a CIE plasma, on the other hand, origins of the line emissions are more 
dominated by the collisional excitation. Therefore, a $kT_z$ determination 
done by comparing with a CIE plasma code is not a proper method. 
Also, the previous works assumed that the continuum spectrum purely consists 
of bremsstrahlung emission, and determined $kT_e$ to be 
$\sim 1.0$~keV. This value has been reduced owing to 
the discovery of the strong RRC emissions.

It should be noted that the elemental abundances in Table~\ref{tab:fit}
are determined as the parameters of the 0.6~keV VAPEC (CIE) component,
and hence should be modified in the real case of the overionized plasma.
The intensity of the Si-He$\alpha$ line is given as
$F_{{\rm He}\alpha} \propto
\varepsilon (T_e) \cdot f_{\rm He}(T_z) \cdot Z_{\rm Si}$,
where $\varepsilon (T_e)$ and $f_{\rm He}(T_z)$ are, respectively,
emissivity coefficient for electron temperature $T_e$ and
fraction of He-like ion for ionization temperature $T_z$.
The abundance, therefore, can be modified by the fraction ratio of
He-like ions, $f_{\rm He}(0.6~{\rm keV})/f_{\rm He}(1.0~{\rm keV})$.
According to the ionization population calculations by 
Mazzotta et al.\ (1998), the ion fractions of He-like,
H-like, and fully-ionized Si ($f_{\rm He}$, $f_{\rm H}$, $f_0$) are
estimated to be (0.83, 0.15, 0.01) for $kT_z$ = 0.6~keV and
(0.37, 0.43, 0.19) for $kT_z$ = 1.0~keV, respectively. Then, the 
modified $Z_{\rm Si}$ is $0.82 \times (0.83/0.37) \simeq 2.2$~solar.
For S, ($f_{\rm He}$, $f_{\rm H}$, $f_0$) are (0.91, 0.02, 0) for 
$kT_z$ = 0.6~keV and (0.59, 0.32, 0.06) for $kT_z$ = 1.2~keV. 
Thus, we similarly modified $Z_{\rm S}$ to be 
$1.7 \times (0.91/0.59) \simeq 2.6$~solar.

Using the measured RRC flux, we can independently determine the volume 
emission measure, VEM = $\int n_e n_p~dV/(4\pi D^2)$, from a following 
equation: 
\begin{equation}
  F_{\rm RRC} = \alpha _1(T_e) \cdot 
  n_Z/n_p \cdot f_0 \cdot {\rm VEM}~,
\label{eq:vem}
\end{equation}
where $\alpha _1(T_e)$ and $n_Z$ are K-shell RRC rate coefficient for 
electron temperature $T_e$ and number density of element $Z$, respectively. 
According to Badnell~(2006), we find total radiative recombination rate 
for fully-ionized Si, at $kT_e$ = 0.6~keV, is 
$\alpha _{\rm tot}$ $\sim$ $2.3 \times 10^{-12}$~cm$^3$~s$^{-1}$. 
Since the rate of the recombination into a level of principal quantum 
number $n$ is described as; 
\begin{equation}
  \alpha _n \propto \frac{1}{n^3}
  \left( \frac{3}{2} \frac{kT_e}{I_z} + \frac{1}{n^2} \right)^{-1} 
  \label{eq:rate}
\end{equation}
(e.g., Nakayama \& Masai 2001), 
we obtain $\alpha _1 / \alpha _{\rm tot}$ = 0.65, for $kT_e$ = 0.6~keV 
and $I_z$ = 2.666~keV. Thus, the VEM is calculated to be 
$\sim 9.9 \times 10^{12}$ 
$(Z_{\rm Si}/2.2~{\rm solar})^{-1}$ $(f_0/0.19)^{-1}$~cm$^{-5}$. 
Similarly, $\alpha _{\rm tot}$ at 0.6~keV for fully-ionized S is 
derived to be $\sim 3.2 \times 10^{-12}$~cm$^3$~s$^{-1}$. Therefore, 
the RRC flux of S corresponds to the VEM of $\sim 9.4 \times 10^{12}$ 
$(Z_{\rm S}/2.6~{\rm solar})^{-1}$ $(f_0/0.06)^{-1}$~cm$^{-5}$. 
These values are almost consistent with that of the VAPEC component 
(Table~\ref{tab:fit}).

The rectangular region in Figure~\ref{fig:image} corresponds to 
$3.1 \times 2.2$~pc$^2$ at $D$ = 1.5~kpc. Assuming the plasma depth of 
3~pc, the emitting volume is estimated to be $6.1 \times 10^{56}$~cm$^3$. 
Therefore, the VEM of $6.6 \times 10^{12}$~cm$^{-5}$ is converted to  
the uniform electron density of $n_e \sim 1.7$~cm$^{-3}$.

As the mechanism to form the overionized plasma, Kawasaki et al.\ (2002) 
proposed that the SNR consists of a central hot ($kT_e$ $\sim$ 1.0~keV) 
region surrounded by a cool ($kT_e$ $\sim$ 0.2~keV) outer shell, and 
interpreted that the hot interior cooled down via thermal conduction to 
the cool exterior. Under the several reasonable boundary conditions, 
they estimated the cooling time from $kT_e$ = 1.5~keV to 1.0~keV is 
about (3--10) $\times$ $10^3$~yr, roughly the same as the SNR age 
($\sim$4000~yr: Troja et al.\ 2008). 
However, this scenario cannot work on our new results. 
According to equation~(5) of Kawasaki et al.\ (2002), 
the conduction timescale is estimated to be 
\begin{equation}
  t_{\rm cond} \simeq 3.9 \times 10^4 
  \left( \frac{n_e}{1.7~{\rm cm}^{-3}} \right) 
  \left( \frac{kT_e}{0.6~{\rm keV}} \right)^{-5/2} {\rm yr}~,  
  \label{eq:conduction}
\end{equation}
if the similar boundary condition is assumed. 
Thus, cooling via conduction requires far longer time than the SNR age. 
Photo-ionization is also unlikely because no strong ionizing source 
is found. Furthermore, the temperature of $\sim$0.6~keV is significantly 
higher than that of a typical photo-ionization plasma 
($\lesssim$0.1~keV: e.g., Kawashima \& Kitamoto 1996).

Since the progenitor of IC~443 has been suggested to be a massive star with 
strong stellar wind activity (Braun \& Strom 1986; Meaburn et al.\ 1990), 
we propose another possibility that the rapid and drastic cooling is 
due to a rarefaction process, as discussed by Itoh \& Masai (1989). 
If a supernova explodes in a dense circumstellar medium made in 
the progenitor's super giant phase, the gas is shock-heated to high 
temperature and significantly ionized at the initial phase of the SNR 
evolution. Subsequent outbreak of the blast wave to a low-density ISM 
caused drastic adiabatic expansion of the shocked gas and resultant 
rapid cooling of the electrons.  
The lifetimes of the fully-stripped ions are roughly estimated to be 
$\tau$ = $(\alpha _{\rm tot} n_e)^{-1}$ 
$\simeq 8.1 \times 10^3 (n_e/1.7{\rm cm}^{-3})^{-1}$~yr for Si and 
$\simeq 5.8 \times 10^3 (n_e/1.7{\rm cm}^{-3})^{-1}$~yr for S, 
respectively. Note that these values are underestimated compared to the 
actual timescale for the plasma to reach CIE, because the contribution of 
collisional ionization processes is ignored. Nevertheless, the estimated 
lifetimes are longer than the age of IC~443. The overionized plasma can, 
therefore, still survive at present.

Future observations with very high energy resolution like the {\it Astro-H} 
mission will give firm evidence for the RRC structure not only on Si and S 
but also on the other major elements. 
This will provide more quantitative study on the peculiar SNR IC~443.

\acknowledgments

The authors deeply appreciate the referee, Randall Smith, for his 
constructive suggestions on revising manuscript. 
We also acknowledge helpful discussions with Kazuo Makishima and Aya Bamba. 
H.\ Yamaguchi and J.\ S.\ Hiraga are supported by the Special Postdoctoral 
Researchers Program in RIKEN. M.\ Ozawa is a Research Fellow of Japan 
Society for the Promotion of Science (JSPS). 
This work is partially supported by the Grant-in-Aid for the Global COE 
Program "The Next Generation of Physics, Spun from Universality and 
Emergence", Young Scientists (HY), and Challenging Exploratory Research (KK) 
from the Ministry of Education, Culture, Sports, Science and Technology (MEXT) 
of Japan.


\end{document}